\documentclass[lettersize,journal]{IEEEtran}
\usepackage{amsmath,amssymb,amsfonts}
\usepackage{algorithmic}
\usepackage{algorithm}
\usepackage{array}
\usepackage{booktabs}
\usepackage{multirow}
\usepackage{textcomp}
\usepackage{stfloats}
\usepackage{url}
\usepackage{verbatim}
\usepackage{graphicx}
\usepackage{cite}
\usepackage{cleveref}
\graphicspath{{figures/}}
\begin{document}

%===============================================================================
% TITLE AND AUTHORS (from micro; formatted for IEEEtran)
%===============================================================================
\title{Trilinear Compute-in-Memory Architecture for Energy-Efficient Transformer Acceleration}

\author{Md~Zesun~Ahmed~Mia,
        Jiahui~Duan,
        Kai~Ni,
        and~Abhronil~Sengupta%
\thanks{M.~Z.~A.~Mia and A.~Sengupta are with the Department of Electrical Engineering,
The Pennsylvania State University, University Park, PA 16802 USA
(e-mail: zesun.ahmed@psu.edu; sengupta@psu.edu).}%
\thanks{J.~Duan and K.~Ni are with the Department of Electrical Engineering,
University of Notre Dame, Notre Dame, IN 46556 USA
(e-mail: jduan3@nd.edu; kni@nd.edu).}%
%\thanks{Manuscript received XXXX XX, 202X; revised XXXX XX, 202X.}
}

%\markboth{IEEE Transactions on Computer-Aided Design of Integrated Circuits and Systems,~Vol.~XX, No.~XX, XXXX~202X}%
%{Mia \MakeLowercase{\textit{et al.}}: Trilinear Compute-in-Memory for Transformer Acceleration}

\maketitle

\begin{abstract}
% ~150 words | Structure: Problem → Gap → Approach → Accuracy → PPA → Significance

Self-attention in Transformers generates dynamic operands that force conventional Compute-in-Memory (CIM) accelerators into costly non-volatile memory (NVM) reprogramming cycles, degrading throughput and stressing device endurance. Existing solutions either reduce but retain NVM writes through matrix decomposition or sparsity, or move attention computation to digital CMOS at the expense of NVM density. We present TrilinearCIM, a Double-Gate FeFET (DG-FeFET)-based architecture that uses back-gate modulation to realize a three-operand multiply-accumulate primitive for in-memory attention computation without dynamic ferroelectric reprogramming. Evaluated on BERT-base (GLUE) and ViT-base (ImageNet and CIFAR), TrilinearCIM outperforms conventional CIM on seven of nine GLUE tasks while achieving up to 46.6\% energy reduction and 20.4\% latency improvement over conventional FeFET CIM at 37.3\% area overhead. To our knowledge, this is the first architecture to perform Transformer attention MAC stages in NVM cores, eliminating runtime dynamic-attention reprogramming.

\end{abstract}

\begin{IEEEkeywords}
Compute-in-memory, transformer acceleration, ferroelectric FET, 
attention mechanism, energy-efficient computing, neuromorphic computing
\end{IEEEkeywords}

%===============================================================================
% I. INTRODUCTION (MERGED from Sections I + II)
% Flow: Problem → Prior Solutions (interleaved) → Gap → Our Solution → Contributions
%===============================================================================
\section{Introduction}
\label{sec:introduction}

% PARA 1: Transformer Revolution & Scaling Success
The Transformer architecture~\cite{vaswani2017attention} has transformed the landscape of deep learning, shifting the paradigm from recurrent and convolutional inductive biases to mechanisms based purely on self-attention. This shift has enabled unprecedented scaling in Natural Language Processing (NLP), with models such as BERT~\cite{devlin2019bert} and GPT-3~\cite{brown2020language} achieving state-of-the-art results by scaling to massive parameter counts (e.g., 175 billion parameters). 
% PARA 2: Vision Expansion & Computational Cost Problem
Following this success in NLP, the Transformer architecture has successfully expanded into Computer Vision with the Vision Transformer (ViT)~\cite{dosovitskiy2020vit} and hierarchical variants like the Swin Transformer~\cite{liu2021swin}, proving that attention-based models can outperform Convolutional Neural Networks (CNNs) at scale. However, this performance comes at a massive computational cost. The quadratic complexity of the self-attention mechanism ($O(N^2)$ with respect to sequence length) combined with massive parameter counts has led to skyrocketing energy consumption for both training and inference, raising significant concerns regarding carbon footprint and sustainability~\cite{strubell2019energy}.

% PARA 3: Algorithmic Optimizations (Prior Solutions - Software/Hardware)
In response to these challenges, the research community has actively pursued algorithmic optimizations to tame the computational overhead. Recent surveys on efficient transformers~\cite{tay2022efficient} detail software-level innovations—ranging from sparse attention patterns and low-rank linear approximations to token pruning strategies—that actively reduce the quadratic complexity and FLOPs count. On the hardware side, dedicated accelerators such as A$^3$~\cite{ham2020a3} exploit token and head sparsity to accelerate attention, while hardware-aware approaches such as HAT search~\cite{wang2020hat} and VAQF co-design~\cite{sun2022vaqf} automate the search for efficient architectures tailored to specific constraints. Yet these approaches remain bounded by the separation of memory and compute in von Neumann architectures.
% PARA 4: Bipartite Barrier (Gap - Architectural Limitation)
At a deeper level, all these optimizations operate within an inherently two-operand computational model---one that maps only two operands (input and weight) per operation. The self-attention mechanism, by contrast, requires computing products of three dynamically-generated matrices ($Q$, $K$, $V$), none of which are static weights. Standard hardware lacks a physical ``third operand'' pathway to support this three-way interaction natively, creating a structural mismatch between the two-operand hardware primitive and the dynamic multi-operand dataflow required by self-attention.

% PARA 5: CIM Promise (Prior Solutions - Hardware)
To overcome the von Neumann bottleneck, Compute-in-Memory (CIM) has emerged as a hardware paradigm that eliminates costly weight data movement. By leveraging the physical properties of resistive switching devices---such as resistive random-access memory (ReRAM), phase-change memory (PCM), and ferroelectric field-effect transistors (FeFETs)---CIM arrays perform matrix-vector multiplication (MVM) \textit{in-situ} using analog current summation (output current $I = G \cdot V$), where conductance ($G$) encodes the weight and voltage ($V$) encodes the input~\cite{sebastian2020memory, ielmini2018memory}. Pioneering architectures such as PRIME~\cite{chi2016prime}, ISAAC~\cite{shafiee2016isaac}, and PUMA~\cite{ankit2019puma} demonstrated orders-of-magnitude energy-delay product improvements for CNN workloads.
% PARA 6: Dynamic Attention Problem (Gap - CIM Limitation)
However, CIM relies on a critical assumption: \textit{weight stationarity}. In CNNs and standard multi-layer perceptrons (MLPs), weight matrices are static after training, allowing CIM accelerators to program the non-volatile memory once and reuse it for millions of inference cycles. Transformers violate this assumption: the Self-Attention mechanism generates key operands ($Q, K, V$) dynamically from every input, forcing a ``Compute-Write-Compute'' cycle that repeatedly reprograms NVM arrays. Because NVM writes are orders of magnitude slower, more energy-intensive, and endurance-limited compared to reads~\cite{peng2019neurosim, zhang2024watt}, this reprogramming dominates execution time and degrades device lifetime---a bottleneck we quantify in Section~\ref{sec:motivation}. Architectural solutions such as ReTransformer~\cite{yang2020retransformer} and TransPIM~\cite{zhou2022transpim} have attempted to mitigate the bottleneck through matrix decomposition and dataflow optimization, but cannot eliminate the writes entirely.

% PARA 7: Hybrid Compromise (Gap - Current SOTA Limitation)
Recognizing the insurmountable endurance and latency penalties of NVM writes, recent state-of-the-art accelerators have pivoted towards hybrid architectures. A prime example is X-Former~\cite{sridharan2023xformer}, which partitions the workload: it uses NVM tiles for the static projection weights (where CIM excels) but retreats to digital CMOS engines for the attention mechanism ($Q \cdot K^T$) to avoid writing to NVM. While this hybrid approach effectively solves the endurance problem, it represents a significant retreat from the promise of ``All-in-Memory'' computing. By reverting to CMOS for the attention score computation ($Q \cdot K^T$, which scales quadratically with sequence length), these designs sacrifice the superior area density and leakage characteristics of NVM—accepting a trade-off: \textit{solve endurance by sacrificing area/efficiency}.

% PARA 8: Our Solution - TrilinearCIM
To resolve this dichotomy, we propose \textbf{TrilinearCIM}, an architecture that extends the conventional FeFET-based CIM model. A ferroelectric field-effect transistor (FeFET) stores a non-volatile conductance state in its ferroelectric gate stack, making it well suited for weight-stationary CIM. However, a standard single-gate FeFET still exposes only the usual two operands: stored conductance and applied input voltage. We therefore adopt a double-gate FeFET (DG-FeFET) structure~\cite{mulaosmanovic2021ferroelectric, jiang2022asymmetric, jiang2025bio}, in which a ferroelectric top gate stores the non-volatile weight while a non-ferroelectric back gate provides an independent volatile modulation path. This additional control terminal creates the third operand pathway needed by TrilinearCIM to execute attention MAC stages in-memory while eliminating runtime dynamic-attention reprogramming.
% CONTRIBUTIONS
Our specific contributions are as follows.

\textbf{(1) Novel Trilinear CIM Primitive.} We propose a three-operand Compute-in-Memory operation ($Y = A \cdot B \cdot C$) enabled by the electrical properties of the DG-FeFET back-gate. This extends the standard two-operand ($\text{Input} \cdot \text{Weight}$) CIM paradigm to natively support three-operand operations.

\textbf{(2) Elimination of Dynamic NVM Writes.} By mapping static weights ($W_Q, W_K, W_V$) to the non-volatile top gate and dynamic activation vectors ($X, X^T$) to the volatile back-gate voltage ($V_{BG}$), TrilinearCIM computes the attention mechanism within the memory array without reprogramming the ferroelectric state for dynamic attention operands.

\textbf{(3) Simplified Dataflow and Reduced Buffer Pressure.} Fusing the projection and attention steps into trilinear stages reshapes the dataflow. The intermediate dynamic matrices are never stored in the global buffer, reducing the number of matrices that must reside simultaneously in the buffer from three ($X$, $Q$, $K$) to one ($X$), thereby lowering buffer capacity requirements by approximately $3\times$.

\textbf{(4) Recovery of Compute Efficiency.} By removing dynamic write overhead from the critical path, our evaluation shows that TrilinearCIM significantly reduces both energy consumption and latency compared to conventional CIM baselines. This restores the competitive advantage of ``All-in-Memory'' computing for Transformer workloads.

%===============================================================================
% II. BACKGROUND
%===============================================================================
\section{Background}
\label{sec:background}

\subsection{Transformer Architecture and Self-Attention}
\label{subsec:transformer_math}
\begin{figure}[t]
    \centering
    \includegraphics[width=0.55\columnwidth]{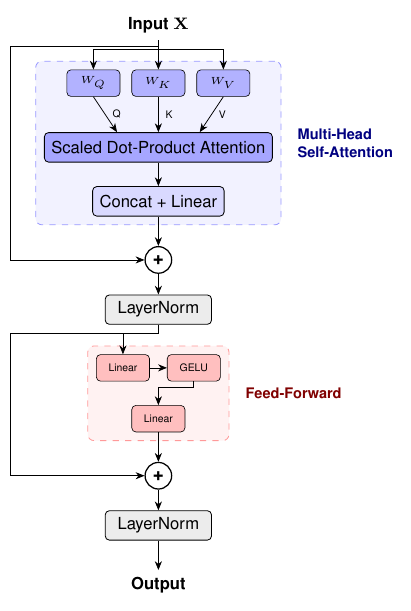}
    \caption{Transformer Encoder Block architecture. The input sequence $X$ is projected through learned weight matrices $W_Q$, $W_K$, $W_V$ to produce Query ($Q$), Key ($K$), and Value ($V$) representations. These are processed by Scaled Dot-Product Attention (Eq.~\ref{eq:attention}), followed by concatenation and linear projection to form the Multi-Head Self-Attention output. The Feed-Forward Network applies two linear transformations with GELU activation. Residual connections (circled ``+'') add the sub-layer input to its output before Layer Normalization (LayerNorm).}
    \label{fig:transformer_block}
\end{figure}

The standard Transformer architecture is composed of stacked encoder blocks, where each block processes an input sequence $X \in \mathbb{R}^{N \times d}$ (with $N$ tokens and embedding dimension $d$) through two sub-layers: Multi-Head Self-Attention (MHSA) and a Feed-Forward Network (FFN), as illustrated in Fig.~\ref{fig:transformer_block}. Both sub-layers employ residual connections (denoted by circled ``+'' symbols) and Layer Normalization (LN), resulting in the following dataflow:
\begin{align}
    Z &= \text{LN}(X + \text{MHSA}(X)) \label{eq:transformer_mhsa_block} \\
    Y &= \text{LN}(Z + \text{FFN}(Z)) \label{eq:transformer_ffn_block}
\end{align}
While the FFN (Eq.~\ref{eq:transformer_ffn_block}) relies purely on static weight matrices that are inherently compatible with standard CIM, the MHSA (Eq.~\ref{eq:transformer_mhsa_block}) introduces dynamic data dependencies.

Mathematically, the attention mechanism first projects the input into Query ($Q$), Key ($K$), and Value ($V$) subspaces as shown in Eq.~\ref{eq:projections}:
\begin{equation}
    Q = X W_Q^T, \quad K = X W_K^T, \quad V = X W_V^T
    \label{eq:projections}
\end{equation}
where $W_Q, W_K, W_V \in \mathbb{R}^{d_k \times d}$ are learnable parameters, and $d_k$ is the dimensionality of the key subspace. The core attention scores are then computed via a scaled dot-product (Eq.~\ref{eq:attention}):
\begin{equation}
    \text{Attention}(Q, K, V) = \text{softmax}\left(\frac{Q K^T}{\sqrt{d_k}}\right) V
    \label{eq:attention}
\end{equation}
Here, the matrix multiplication $Q K^T$ in Eq.~\ref{eq:attention} represents the token-to-token correlation. Unlike the projection steps where weights are static, both operands in this equation differ for every input sample. Finally, Multi-Head Attention aggregates outputs from $h$ parallel heads, each operating on a $d_k$-dimensional subspace:
\begin{equation}
    \text{MultiHead}(Q, K, V) = \text{Concat}(\text{head}_1, \dots, \text{head}_h)W_O
    \label{eq:multihead}
\end{equation}
where $W_O \in \mathbb{R}^{d \times d}$ is the output projection matrix. As defined in Eq.~\ref{eq:multihead}, this formulation underscores the distinct requirements of the two sub-layers: static weights for projections/FFNs, and dynamic variable-variable multiplication for attention scores.

\subsection{CIM and the DG-FeFET Device}
\label{subsec:cim_dgfefet}

% TRANSITION from 2.1 - connects to "static vs dynamic" conclusion
To execute these matrix operations efficiently, Compute-in-Memory (CIM) arrays exploit analog physics. In a resistive memory array, the current through each device, $I$, follows Ohm's Law ($I = G \cdot V$), where $G$ is the programmable conductance (representing a weight) and $V$ is the applied voltage (representing an input). By summing currents along the columns via Kirchhoff's Current Law, the array computes the dot product in a single step:
\begin{equation}
    I_j = \sum_{i} G_{ij} \cdot V_i
    \label{eq:cim_mvm}
\end{equation}
where $G_{ij}$ is the cell conductance at row $i$ and column $j$, and $V_i$ is the input voltage applied to row $i$. The resulting current $I_j$ collected along column $j$ represents the dot-product output. This eliminates the repeated movement of stored weights from memory to compute units, which dominates energy consumption in conventional von Neumann systems.

\begin{figure}[t]
    \centering
    \includegraphics[width=0.95\linewidth]{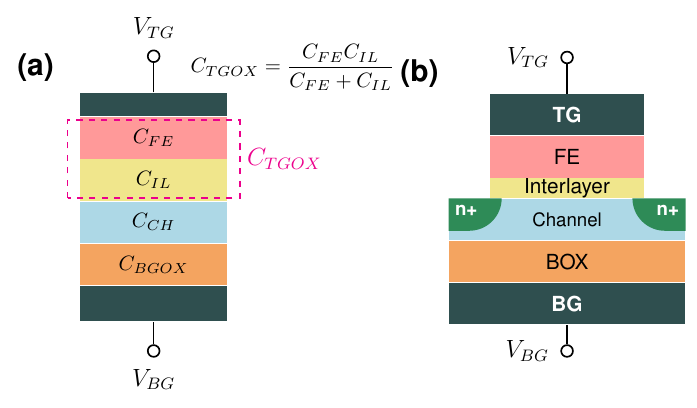}
    \caption{DG-FeFET device concept. (a)~Capacitor stack model illustrating the series combination of ferroelectric ($C_{FE}$) and interlayer ($C_{IL}$) capacitances forming the effective top-gate oxide capacitance $C_{TGOX}$. (b)~Device cross-section showing the Top-Gate (TG), Back-Gate (BG), and localized n$^{+}$ source/drain implant regions within the channel.}
    \label{fig:dgfefet_concept}
\end{figure}

% DG-FeFET INTRODUCTION - addresses 3-operand need
However, standard CIM devices still support only \textit{two} operands per operation. To support the trilinear computations used in our attention dataflow ($A \cdot B \cdot C$), we leverage the DG-FeFET device introduced in Section~\ref{sec:introduction}, whose back-gate provides a volatile third-operand pathway.
% DEVICE STRUCTURE
As illustrated in Fig.~\ref{fig:dgfefet_concept}(b), this device integrates two electrically-coupled gates: a \textbf{Top-Gate~(TG)}, whose ferroelectric layer stores the \textit{non-volatile} weight via polarization state, and a \textbf{Back-Gate~(BG)}, a volatile control terminal separated by a standard dielectric (buried oxide) that provides dynamic modulation of the channel conductance. The thin, fully-depleted silicon channel is sandwiched between these gates, enabling strong electrostatic coupling.

% PHYSICS: Vth SHIFT
Physically, the back-gate voltage ($V_{BG}$) linearly modulates the effective threshold voltage ($V_{th}$) of the top-gate~\cite{mulaosmanovic2021ferroelectric, lim1983threshold}. This coupling is characterized by the coefficient $\gamma_{TG}$, defined by the equivalent capacitance network shown in Fig.~\ref{fig:dgfefet_concept}(a):
\begin{equation}
    \gamma_{TG} = \frac{C_{CH} \cdot C_{BGOX}}{C_{TGOX}(C_{CH}+C_{BGOX})}
    \label{eq:gamma_tg}
\end{equation}
where $C_{CH}$ and $C_{BGOX}$ are the channel and back-gate oxide capacitances. The effective top-gate capacitance $C_{TGOX}$ is the series combination of the ferroelectric capacitance $C_{FE}$ and the interfacial layer capacitance $C_{IL}$:
\begin{equation}
    C_{TGOX} = \frac{C_{FE} \cdot C_{IL}}{C_{FE} + C_{IL}}
    \label{eq:ctgox}
\end{equation}
The resulting threshold voltage shift is:
\begin{equation}
    \Delta V_{th} = -\gamma_{TG} \cdot V_{BG}
    \label{eq:vth_shift}
\end{equation}

% PHYSICS: CONDUCTANCE MODULATION (EXPANDED)
This threshold shift, combined with mobility enhancement at higher $V_{BG}$ (the carrier centroid shifts away from the top-gate interface, reducing Coulomb and surface roughness scattering~\cite{nier2013multi, al2022impact, han2022back}), translates into a \textit{multiplicative} modulation of the channel conductance $G_{DS}$. In the deep triode regime, where $\mu_n(V_{BG})$ denotes the field-dependent electron mobility, the full relationship is~\cite{jiang2025bio}:
\begin{equation}
    G_{DS}(V_{BG}) = \frac{\mu_n(V_{BG})}{\mu_n(0)} \cdot G_{DS}(0) + \gamma_{TG} \cdot \mu_n(V_{BG}) \cdot C_{TGOX} \cdot V_{BG}
    \label{eq:gds_full}
\end{equation}
where $G_{DS}(0)$ is the channel conductance at zero back-gate bias (hereafter denoted $G_0$). Consistent with the linear $G_{DS}(V_{BG})$ behavior reported for DG-FeFETs~\cite{jiang2025bio}, the mobility can be approximated as $\mu_n(V_{BG}) \approx \mu_n(0)(1 + \alpha V_{BG})$, where $\alpha$ is the mobility-sensitivity coefficient~\cite{nier2013multi}. Under this first-order approximation, the conductance response can be written as
\begin{equation}
    G_{DS}(V_{BG}) \approx G_0 \cdot (1 + \eta_{BG} \cdot V_{BG})
    \label{eq:gds_simplified}
\end{equation}
This linearization drops the second-order term $\gamma_{TG}\mu_n(0)\alpha C_{TGOX}V_{BG}^2$. Accordingly, Eq.~\ref{eq:gds_simplified} serves as a first-order approximation for CIM analysis within this linear operating regime.
In the CIM paradigm, $G_0$ encodes the stored weight: it is programmed once during model initialization and remains stationary throughout inference, serving as the stationary weight operand of the trilinear primitive developed in Section~\ref{sec:architecture}. Comparing Eqs.~\ref{eq:gds_full} and~\ref{eq:gds_simplified}, and defining the electrostatic coupling coefficient $M = \gamma_{TG} \cdot C_{TGOX} \cdot \mu_n(0)$, we identify the modulation sensitivity:
\begin{equation}
    \eta_{BG} = \alpha + \frac{M}{G_0}
    \label{eq:eta_bg_derivation}
\end{equation}
The first term ($\alpha$) arises from the linear mobility dependence on $V_{BG}$ and captures the mobility-enhancement contribution, while the second term $M/G_0$ captures electrostatic threshold modulation; $\eta_{BG}$ therefore represents the combined first-order back-gate sensitivity. We extract $\alpha = 0.137$~V$^{-1}$ and $M = 1.54$~$\mu$S/V by numerically fitting our physics-inspired polynomial constraints to the experimentally reported $G_{DS}$ vs.~$V_{BG}$ data from Jiang et al.~\cite{jiang2025bio}. This fit provides the experimentally calibrated DG-FeFET back-gate modulation model used later in our trilinear-array evaluation.
% TRILINEAR CAPABILITY - CONCLUSION
Eq.~\ref{eq:gds_simplified} captures the key device behavior for this work: the back-gate voltage modulates the stored conductance multiplicatively through the effective sensitivity $\eta_{BG}$. This provides the physical basis for the trilinear CIM primitive developed in Section~\ref{subsec:trilinear}.

%===============================================================================
% III. MOTIVATION ANALYSIS
%===============================================================================
\section{Motivation: Why Trilinear CIM?}
\label{sec:motivation}

Having established the device physics of the DG-FeFET, we now quantify the cost of the conventional write-based approach to attention.

\subsection{Write Bottleneck and Endurance}
\label{subsec:write_bottleneck}

The ``Compute-Write-Compute'' dataflow introduced in Section~\ref{sec:introduction} is severely penalized by the read/write asymmetry intrinsic to NVM devices. Table~\ref{tab:read_write_asymmetry} quantifies this asymmetry for FeFET devices.

\begin{table}[h]
\centering
\caption{FeFET read vs. write asymmetry~\cite{jerry2017fefet, peng2019neurosim}.}
\label{tab:read_write_asymmetry}
\begin{tabular}{lcc}
\toprule
\textbf{Metric} & \textbf{Read} & \textbf{Write} \\
\midrule
Latency & $\sim$10 ns & $\sim$50 ns \\
Energy/cell & $\sim$fJ & $\sim$sub-pJ \\
\bottomrule
\end{tabular}
\end{table}

For a BERT-Base configuration ($N = 512$ tokens, $d_k = 64$, $h = 12$ heads, $L = 12$ layers), the aggregate runtime programming volume becomes:
\begin{equation}
    N_{\mathrm{prog}} = 2 \cdot N \cdot d_k \cdot h \cdot L \cdot \left\lceil \frac{8}{2} \right\rceil \cdot 2 \approx 75.5\text{M}
    \label{eq:write_count}
\end{equation}
where the first factor of 2 accounts for the two dynamic operands ($K^T$ and $V$), $\lceil 8/2 \rceil = 4$ maps each 8-bit value onto four 2-bit FeFET cells, and the final factor of 2 reflects separate positive and negative arrays for signed representation. This write volume has severe implications for device lifetime. FeFET devices exhibit $10^{6}$--$10^{12}$ write cycles depending on oxide quality~\cite{jerry2017fefet}. Because the temporary attention arrays must be reprogrammed during every inference, these runtime writes repeatedly stress the cells assigned to $K^T$ and $V$ storage. Moreover, this estimate is for BERT-Base---a relatively small model. Scaling to BERT-Large ($h{=}16$, $L{=}24$) would increase the aggregate programming volume by approximately 2.7$\times$. Even technologies with substantially higher nominal endurance, such as STT-MRAM ($>$$10^{12}$~\cite{chen2016nvm}) or SOT-MRAM ($>$$10^{15}$~\cite{vanbeek2023sot}), would still face the same fundamental bottleneck: runtime rewriting of attention operands grows with model size and sequence length. Together, the latency, energy, and endurance penalties of this repeated reprogramming critically limit the viability of write-based CIM for dynamic attention.

\subsection{Case for Trilinear Operations}
\label{subsec:trilinear_case}

The DG-FeFET resolves the two-operand limitation by providing a third \textit{volatile} operand pathway through the back-gate (Section~\ref{subsec:cim_dgfefet}). This additional control terminal enables a different attention dataflow: instead of repeatedly reprogramming dynamic operands into non-volatile memory, the trilinear approach keeps projection weights stationary in the array and applies input-dependent modulation through the back-gate. As a result, attention computations can be carried out \textit{in-situ} without runtime ferroelectric rewriting. This directly addresses the two bottlenecks identified in Section~\ref{subsec:write_bottleneck}: \textbf{(i)}~write latency is eliminated because dynamic operands are conveyed through back-gate voltage modulation rather than polarization switching; and \textbf{(ii)}~runtime dynamic-attention reprogramming is eliminated because attention execution conveys temporary operands through volatile back-gate modulation rather than ferroelectric switching.

%===============================================================================
% IV. PROPOSED ARCHITECTURE
%===============================================================================
\section{Proposed Trilinear CIM Architecture}
\label{sec:architecture}

%-------------------------------------------------------------------------------
% 4A: System Overview (NEW)
%-------------------------------------------------------------------------------
\subsection{System Overview}
\label{subsec:system_overview}

\begin{figure}[!t]
    \centering
    \includegraphics[width=0.95\columnwidth]{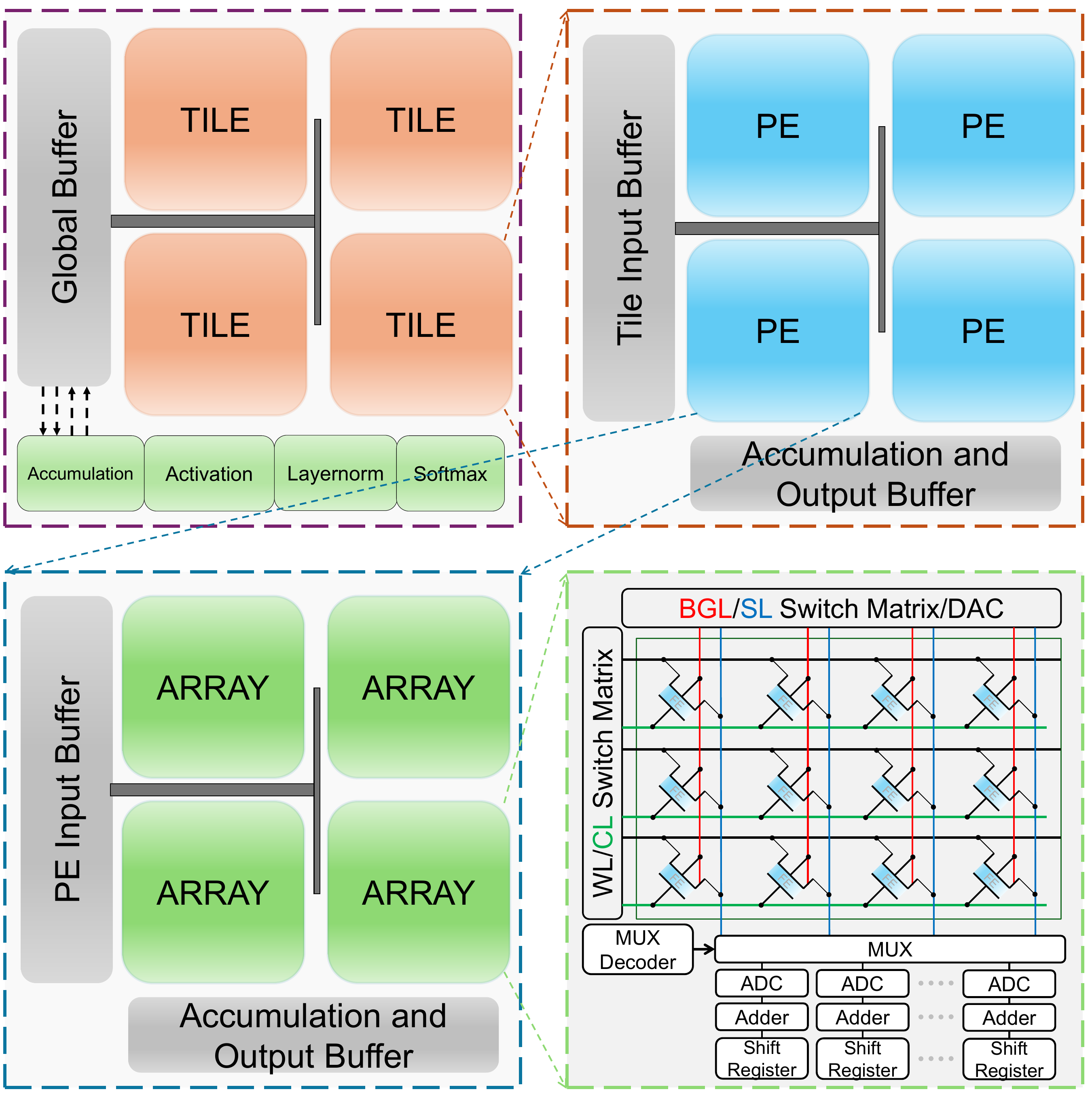}
    \caption{Hierarchical organization of the trilinear CIM accelerator. \textbf{(Top-left)} Chip level: 2$\times$2 tile grid with global buffer, accumulation unit, and Special Function Unit (SFU) for Softmax, LayerNorm, and Activation. \textbf{(Top-right)} Tile level: 2$\times$2 Processing Element (PE) grid with tile input buffer and accumulation/output buffer. \textbf{(Bottom-left)} PE level: 2$\times$2 SubArray grid with PE input buffer. \textbf{(Bottom-right)} SubArray level: DG-FeFET cells driven by wordline/control-line (WL/CL) switch matrix (rows) and back-gate-line/source-line (BGL/SL) switch matrix/DAC (columns), with multiplexer (MUX)$\rightarrow$ADC$\rightarrow$Adder$\rightarrow$Shift Register readout pipeline.}
    \label{fig:chip_hierarchy}
\end{figure}

Fig.~\ref{fig:chip_hierarchy} illustrates our hierarchical DG-FeFET accelerator architecture, which performs attention through static trilinear computation.

\textbf{Chip Level.}
The top-level architecture organizes tiles in a scalable 2D mesh connected via H-tree interconnect for balanced latency~\cite{chen2018neurosim}. A global buffer interfaced with the tile array stores input sequences $X \in \mathbb{R}^{N \times d}$ and broadcasts them to the tiles. Peripheral to the compute fabric, dedicated functional units (Softmax, LayerNorm, Activation) handle operations incompatible with analog computation. This split compute model—analog multiplication for attention scores, digital for non-linearities—maximizes energy efficiency while maintaining accuracy. Grid dimensions (tile array size, processing elements (PEs) per tile, SubArrays per PE) are determined by the same NeuroSim-based floorplanning flow~\cite{peng2019neurosim} used in our TransCIM evaluation framework (Section~\ref{subsec:transcim}), based on model weight dimensions and target chip organization.

\textbf{Tile Level.}
Each tile comprises a grid of PEs with shared local buffering. The tile input buffer retains frequently reused operands and intermediate vectors, reducing global memory traffic. Partial sums from PE outputs converge through an accumulation network before being written to the tile output buffer, supporting parallel execution across attention heads and partitioned embedding dimensions.

\textbf{PE and SubArray Microarchitecture.}
Within each PE, DG-FeFET arrays support the trilinear attention operations described in Section~\ref{subsec:trilinear} using DG-FeFET cells in a selector-less configuration. Row drivers control two signals: wordlines (WL) carry input activations $X$ to device drains, while control lines (CL) bias the top-gate---enabling row selection during inference or applying programming voltage ($V_{TG}$) during weight updates. Column-wise drivers handle back-gate lines (BGL) that apply the dynamic modulator voltage ($V_{BG}$) via integrated DACs, and source lines (SL) that collect output currents. The resulting analog currents flow to a multi-stage readout pipeline: column multiplexer for output selection, time-multiplexed shared ADCs for digitization, digital adders for partial sum accumulation, and shift registers for multi-bit weight alignment. This mixed-signal readout organization balances throughput with precision by amortizing ADC resources across columns via time-multiplexing. Static linear layers outside attention---including the FFN projections and output projection layers---are mapped to separate single-gate FeFET CIM arrays, whereas the DG-FeFET back-gate is exercised only for attention stages that require dynamic modulation.

%-------------------------------------------------------------------------------
% 4B: Trilinear Operation Concept (formerly 4A)
%-------------------------------------------------------------------------------
\subsection{Trilinear Operation Concept}
\label{subsec:trilinear}

The DG-FeFET enables a trilinear multiply-accumulate (MAC) primitive. From the conductance modulation (Eq.~\ref{eq:gds_simplified}), the full output current is:
\begin{equation}
I_D = V_D \cdot G_0 \cdot (1 + \eta_{BG} \cdot V_{BG})
\label{eq:trilinear_primitive}
\end{equation}
where $V_D$ encodes input activations, $G_0$ stores weight values, and $V_{BG}$ provides dynamic conductance modulation. The trilinear product of interest resides in the second term ($V_D \cdot G_0 \cdot \eta_{BG} \cdot V_{BG}$), while the first term ($V_D \cdot G_0$) is a constant DC bias that is removed through baseline subtraction (Section~\ref{sec:implementation}). Table~\ref{tab:stage_operands} maps these signals to operand roles across attention stages.

\begin{figure}[t]
    \centering
    \includegraphics[width=0.8\columnwidth]{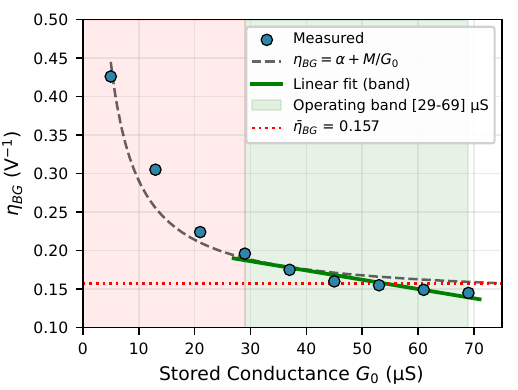}
    \caption{$\eta_{BG}$ vs.\ $G_0$ for the DG-FeFET, following Eq.~\ref{eq:eta_bg_derivation} with parameters $\alpha$ and $M$ extracted from experimental data~\cite{jiang2025bio} (see Section~\ref{subsec:cim_dgfefet}). The green-shaded region indicates the selected operating band ($G_0 \in [29, 69]$~$\mu$S); the dashed horizontal line marks the band-averaged $\bar{\eta}_{BG} = 0.157$~V$^{-1}$ adopted as a uniform constant.}
    \label{fig:eta_bg_operating}
\end{figure}

As derived in Section~\ref{subsec:cim_dgfefet}, $\eta_{BG}$ depends on $G_0$ (Eq.~\ref{eq:eta_bg_derivation}). Fig.~\ref{fig:eta_bg_operating} validates this relationship using the simulated numerical parameters calibrated to the empirical device data~\cite{jiang2025bio}. To ensure uniform trilinear behavior, we constrain $G_0 \in [29, 69]$~$\mu$S---an \textit{operating band} where the residual $\eta_{BG}$ variation remains strictly bounded. Below this range, $\eta_{BG}$ uniformity degrades rapidly, justifying the choice of the lower bound. Within the selected band, we approximate the cell-specific modulation sensitivity $\eta_{BG}$ with a single band-averaged constant, $\bar{\eta}_{BG} = 0.157$~V$^{-1}$. % Accordingly, Eq.~\ref{eq:trilinear_primitive} is evaluated using $\bar{\eta}_{BG}$ in place of the per-cell $\eta_{BG}$. Thus, all cells within the selected conductance band are modeled with the same back-gate response. The remaining cell-to-cell deviations from $\bar{\eta}_{BG}$ introduce only a bounded residual approximation error within the selected operating band.

\subsection{Attention Dataflow}
\label{subsec:dataflow}

Fig.~\ref{fig:dataflow} contrasts conventional two-operand attention with our trilinear approach. In the figure, \textit{Static (Once)} denotes operands associated with one-time programming of non-volatile weights into the array, whereas \textit{Dynamic} denotes inference-time operands that vary with the input sequence. In the conventional dataflow shown in Fig.~\ref{fig:dataflow}(a), query, key, and value projections $Q = X \cdot W_Q^T$, $K = X \cdot W_K^T$, $V = X \cdot W_V^T$ are computed independently, with intermediate tensors stored in and fetched from off-chip DRAM. Once $K$ is available, score computation $Q \cdot K^T$ proceeds, followed by a separate digital scaling step ($\div \sqrt{d_k}$), softmax normalization, and finally $\text{Score} \cdot V$. In a conventional dataflow, softmax and value aggregation can be token-pipelined to hide some latency (i.e., computing the softmax for token $i$ while accumulating the weighted sum over value vectors for token $i-1$). However, this optimization provides diminishing returns because the preceding step---writing the massive intermediate tensors to off-chip DRAM---imposes an overwhelming latency and energy wall that dominates total inference time.

The trilinear dataflow in Fig.~\ref{fig:dataflow}(b) eliminates both the DRAM bottleneck and the separate scaling step by fusing projections with subsequent operations. Like conventional attention, softmax and value aggregation can be token-pipelined; however, trilinear implementation further reduces latency by bypassing explicit $Q$, $K$, and $V$ projection steps entirely---once the static weights are programmed, tokens can stream directly into the attention pipeline: earlier partial results feed downstream stages while later tokens are still being processed, rather than waiting for full tensors to be formed explicitly:

\begin{figure}[t]
    \centering
    \includegraphics[width=\columnwidth]{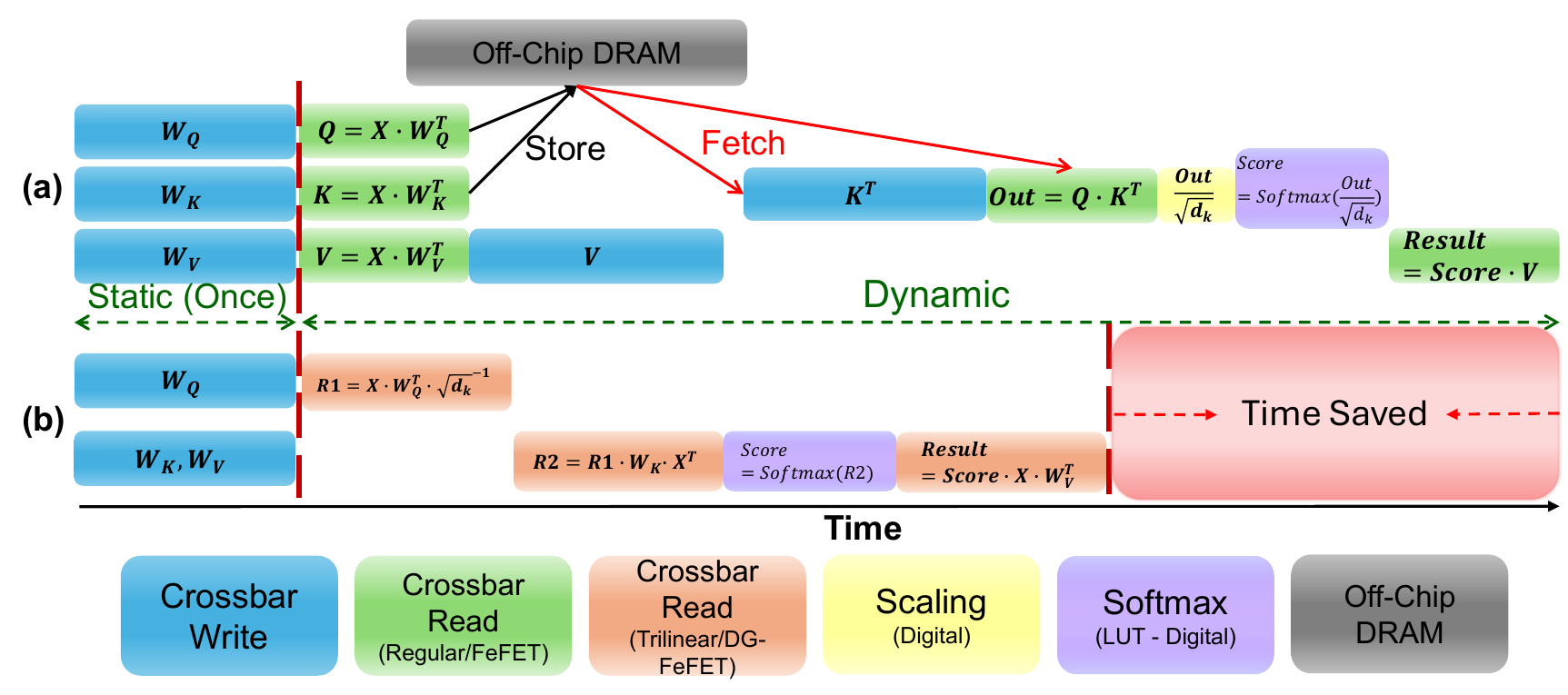}
    \caption{Attention dataflow comparison. (a)~Conventional: $Q/K/V$ projections require off-chip DRAM transfers before score computation. (b)~Trilinear: weights remain static while $X$ modulates device conductance via back-gate, eliminating DRAM traffic and saving compute time.}
    \label{fig:dataflow}
\end{figure}

\textbf{Stage 1: Scaled Query Generation.}
Query vectors with built-in scaling are computed as $R_1 = X \cdot W_Q^T \cdot (1/\sqrt{d_k})$, where $W_Q^T$ is stored in the crossbar, $X$ is applied through the row input path, and the static scaling factor is applied via the back gate. This single trilinear MAC replaces the conventional projection and separate scaling steps.

\textbf{Stage 2: Score Synthesis.}
Attention scores are computed as $R_2 = R_1 \cdot W_K \cdot X^T$. Since $K = XW_K^T$, we have $K^T = W_KX^T$; therefore $QK^T/\sqrt{d_k}$ can be evaluated without forming or writing $K^T$. The query vector $R_1$ is applied as the row-side input, $W_K$ is stored in the crossbar, and $X^T$ modulates conductance dynamically via the back gate. This vector-matrix-vector multiply (VMVM) directly yields pre-softmax scores without computing or storing key tensors.

\textbf{Stage 3: Value Aggregation.}
After digital softmax produces $\text{Score} = \text{softmax}(R_2)$, value aggregation is computed as $\text{Result} = \text{Score} \cdot X \cdot W_V^T$. Since $V = XW_V^T$, this VMVM stage evaluates $\text{Score}\cdot V$ without materializing or writing $V$. The input sequence $X$ provides the row-side input, $\text{Score}$ modulates conductance via back-gate broadcast, and $W_V^T$ is stored.

\textbf{Key Insight: Static vs. Dynamic Modulation.}
Table~\ref{tab:stage_operands} summarizes the operand-to-terminal mapping for each stage, distinguishing fixed from time-varying back-gate modulation during inference. Stage~1 uses \emph{static} modulation: the scaling factor $1/\sqrt{d_k}$ is constant for all tokens and is applied as a fixed back-gate voltage. As a result, this stage does not require dynamic back-gate updates and could equivalently be realized using a standard single-gate FeFET. Stages~2 and~3 use \emph{dynamic} modulation: the back-gate voltage changes with the token-dependent operand ($X^T$ in Stage~2 and $\text{Score}$ in Stage~3). This distinction is architecturally significant because dynamic modulation requires per-token back-gate updates and therefore incurs back-gate delivery overhead.

\textbf{Memory Traffic Reduction.}
The trilinear dataflow completely eliminates the need to store intermediate projection tensors. In conventional attention, a sequence of length $N$ and embedding dimension $d$ requires $3 \times N \times d$ elements of intermediate storage ($Q$, $K$, $V$ projections), which often spills to off-chip DRAM for long sequences. Our approach eliminates the need to store the intermediate projection tensors, while retaining only the input sequence $X$ for reuse and residual connections. This translates to substantial energy savings, as a DRAM access consumes roughly two orders of magnitude more energy than a small on-chip SRAM/cache access~\cite{horowitz20141}.

\begin{table}[t]
\centering
\caption{Trilinear attention stages and operand mapping.}
\label{tab:stage_operands}
\setlength{\tabcolsep}{3pt}
\begin{tabular}{@{}lcccc@{}}
\toprule
\textbf{Stage} & \textbf{Math} & \textbf{Row In.} & \textbf{Stored} & \textbf{BG In.} \\
\midrule
Scaled $Q$ & $X \cdot W_Q^T / \sqrt{d_k}$ & $X$ & $W_Q^T$ & $1/\sqrt{d_k}$ \\
Score & $R_1 \cdot W_K \cdot X^T$ & $R_1$ & $W_K$ & $X^T$ \\
$V$-Agg & $\text{Score} \cdot X \cdot W_V^T$ & $X$ & $W_V^T$ & $\text{Score}$ \\
\bottomrule
\end{tabular}
\end{table}

\subsection{DG-FeFET Crossbar Array Design}
\label{subsec:crossbar}

\begin{figure*}[t]
    \centering
    \includegraphics[width=0.49\textwidth]{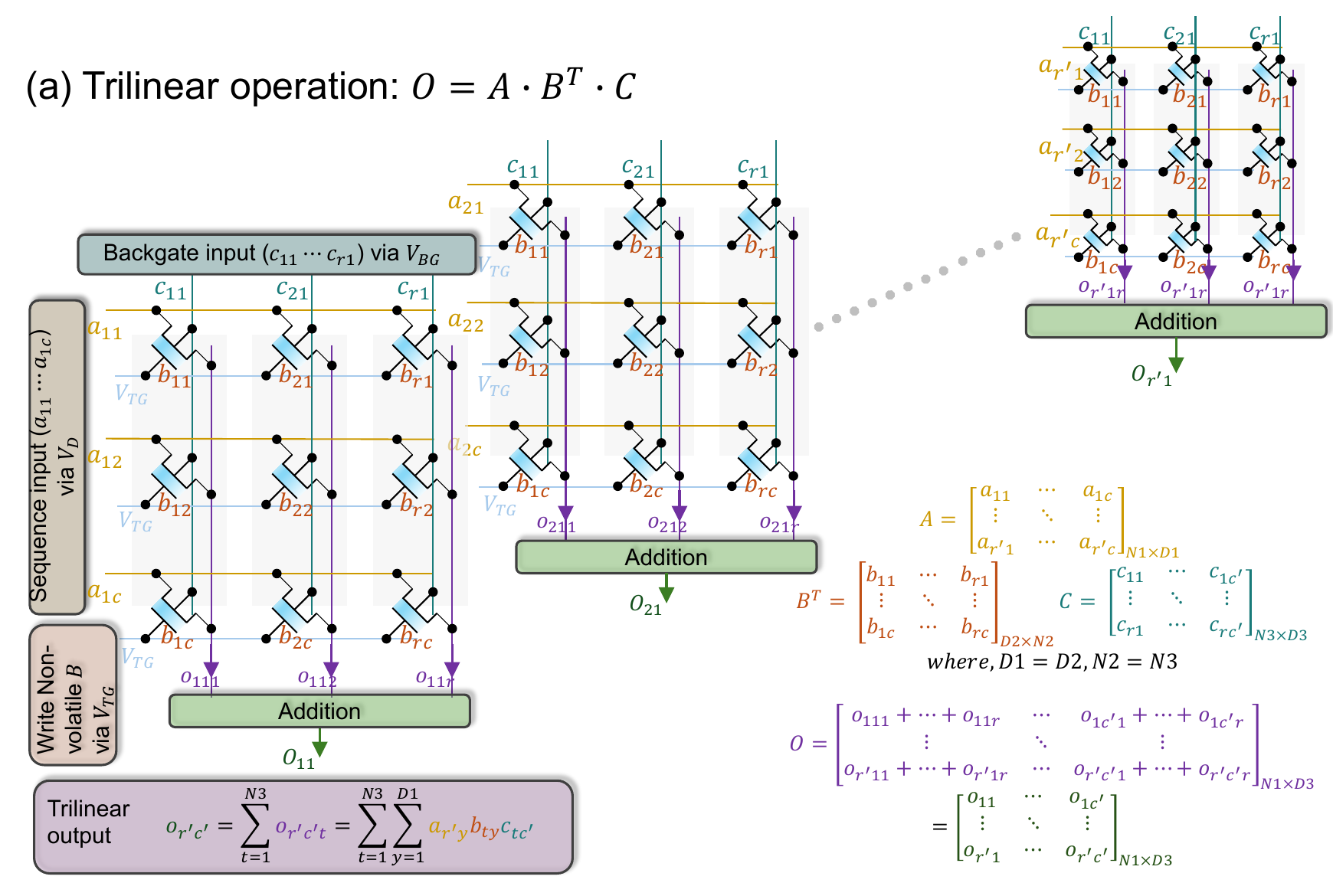}
    \hfill
    \includegraphics[width=0.49\textwidth]{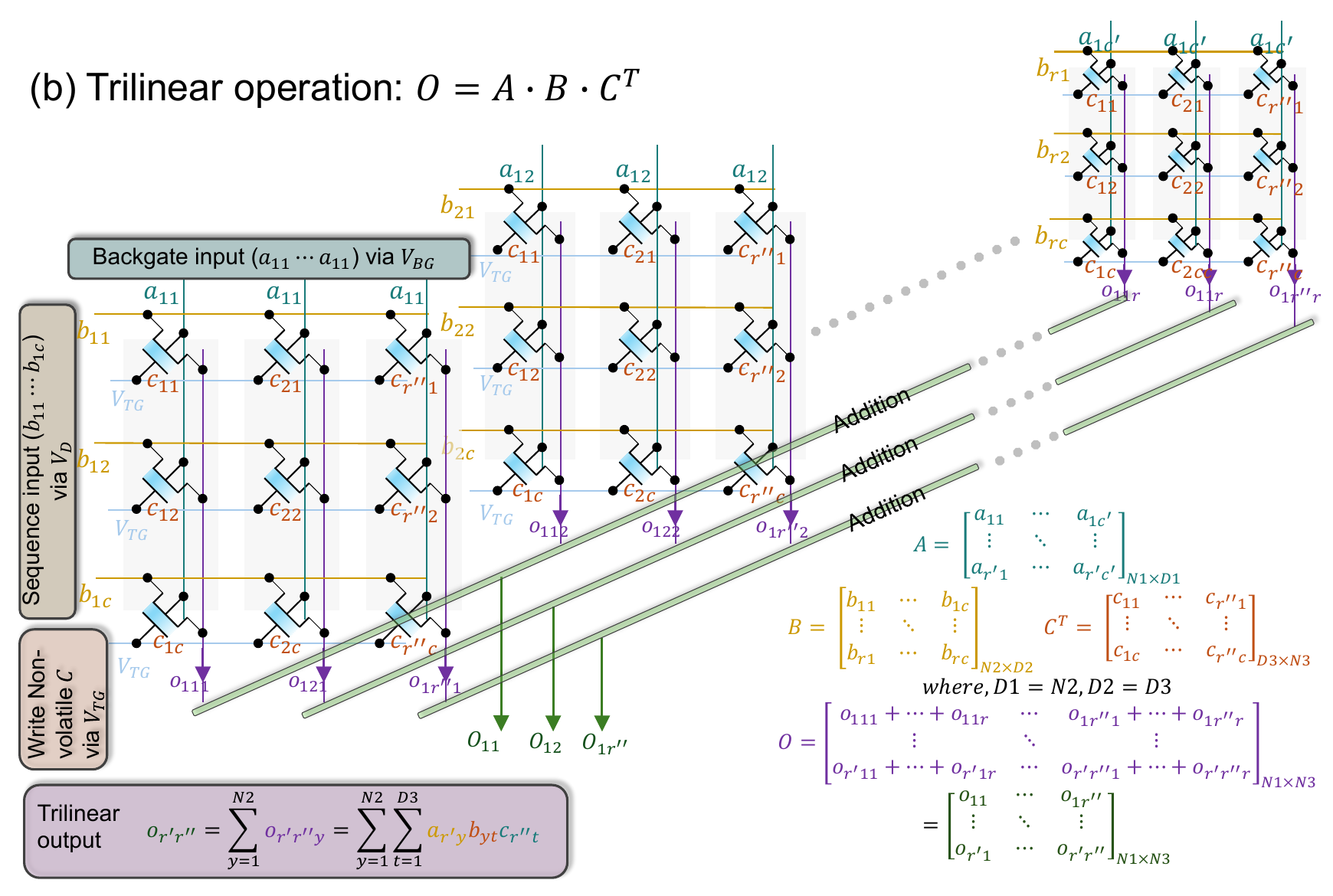}
    \caption{DG-FeFET crossbar array for trilinear MAC. (a)~$O = A \cdot B^T \cdot C$: weights $B^T$ stored in device conductance, input $A$ applied to rows, modulator $C$ applied to columns. (b)~$O = A \cdot B \cdot C^T$: weights $C^T$ stored in device conductance, input $B$ applied to rows, modulator $A$ broadcast across all columns.}
    \label{fig:xbar_trilinear}
\end{figure*}

The DG-FeFET crossbar employs a selector-less architecture where each device connects directly between its row and column lines without an access transistor. The high on/off current ratio ($>10^4$) inherent to FeFETs~\cite{ni2018fefet, yin2018ferroelectric, mulaosmanovic2019ferroelectric} mitigates sneak-path currents. %, achieving 2F$\times$2F cell density (where $F$ is the minimum feature size). 
As illustrated in Fig.~\ref{fig:xbar_trilinear}, the crossbar organization maps directly to the trilinear primitive (Eq.~\ref{eq:trilinear_primitive}): static weights reside in device conductance ($G_0$), sequence inputs stream along rows ($V_D$), and dynamic modulators are applied along columns ($V_{BG}$). The per-column routing enables either independent per-column biasing (for element-wise operand application in Configuration~(a)) or uniform broadcast across all columns (for common-scalar input in Configuration~(b)). 

The two crossbar configurations differ in their accumulation strategy:
\textbf{Configuration~(a) (Fig.~\ref{fig:xbar_trilinear}a): $O = A \cdot B^T \cdot C$.}
Each crossbar stores weight matrix $B^T$ in device conductance. Parallel crossbars compute independent rows of the output matrix over time: crossbar $i$ receives input row $A_{i,:}$ applied to its rows, and back-gate vector $C_{:,j}$ applied to its columns. Within each crossbar, analog column currents sum via Kirchhoff's current law (over the inner-product dimension), and a digital adder aggregates all column outputs. This \textit{intra-crossbar addition} produces one output element $o_{ij}$ per crossbar, per cycle. The back-gate loops through the columns of $C$: in the first cycle, all crossbars receive column $C_{:,1}$ across their back-gates; in the second cycle, $C_{:,2}$; and so on.
\textbf{Configuration~(b) (Fig.~\ref{fig:xbar_trilinear}b): $O = A \cdot B \cdot C^T$.}
Each crossbar stores weight matrix $C^T$ in device conductance---all parallel crossbars share identical weights. Crossbar $i$ takes row $B_{i,:}$ as its row input. Column currents from corresponding positions across all crossbars are summed via \textit{inter-crossbar addition} to produce the output elements. The back-gate loops through the rows of $A$: in the first cycle, crossbar $i$ receives element $a_{1i}$ (broadcast to all its columns); in the second cycle, it receives $a_{2i}$; and so on through $a_{r'i}$.

% \textbf{Temporal Sequencing.}
% In both configurations, the WL input remains fixed while the BGL cycles through all assigned values. After completing the BGL loop, the next WL input row is applied. This pattern minimizes WL switching energy while enabling efficient back-gate modulation.

Stage~2 (Score Synthesis) uses the configuration in Fig.~\ref{fig:xbar_trilinear}(a), mapping $R_1$, $W_K$, and $X^T$ to the row input, stored conductance, and back-gate operand, respectively. Stage~3 (Value Aggregation) uses Fig.~\ref{fig:xbar_trilinear}(b), mapping $X$, $W_V^T$, and $\text{Score}$ to the same three physical roles. Stage~1 (Scaled Query Generation) can use either configuration because its back-gate operand is fixed. This VMVM schedule turns the DG-FeFET third terminal into an attention dataflow and avoids runtime $K^T$ and $V$ writes.

\subsection{Digital Operations Integration}
\label{subsec:digital_ops}

Transformer inference requires several digital operations incompatible with analog memory arrays. These include distributed cross-tile accumulation, as well as dedicated non-linear functions (Softmax, LayerNorm, Activation) handled by a Special Function Unit (SFU) peripheral to the tile array (Fig.~\ref{fig:chip_hierarchy}).

\textbf{Softmax.} Attention score normalization is implemented via a four-stage pipeline: (1)~a comparator tree finds the maximum value across the sequence for numerical stability, (2)~an exponential lookup table (LUT) computes $e^{x_i - x_{\max}}$ for each element (where $x_{\max} = \max_i(x_i)$), (3)~an adder tree sums the exponentiated values, and (4)~a reciprocal LUT followed by multipliers divides each exponential by the sum. The overall pipeline has fixed, deterministic latency, with the LUT stages completing in a single cycle using 256-entry tables for 8-bit precision.

\textbf{LayerNorm.} Layer normalization is implemented as a two-pass pipeline over the $d$-dimensional embedding vector. The first pass computes the mean $\mu$ using an adder tree followed by fixed-point division. The second pass subtracts $\mu$, squares the residuals, accumulates the variance $\sigma^2$, and applies an inverse-square-root LUT to normalize the vector. A final affine stage then applies the learned per-dimension scale and bias to the normalized vector.

\textbf{Activation (GELU).} The FFN sub-layer uses the sigmoid-based approximation $\text{GELU}(x) \approx x \cdot \sigma(1.702\,x)$~\cite{hendrycks2016gelu}, where $\sigma(\cdot)$ denotes the logistic sigmoid function. Our hardware implements this in a three-stage pipeline: (1)~a shift-and-add scaler approximates the constant multiplication $1.702\,x$ without a dedicated multiplier, (2)~a 256-entry sigmoid LUT maps the scaled value to $\sigma(1.702\,x)$, and (3)~a fixed-point multiplier produces the final product $x \cdot \sigma(1.702\,x)$. This decomposition avoids the expensive error-function evaluation of the exact GELU~\cite{hendrycks2016gelu} while reusing the same LUT and multiplier primitives employed by the softmax unit.

\textbf{Accumulation.} Partial sums are reduced via a local hierarchical adder network within each PE and tile. These local accumulation buffers aggregate SubArray and PE outputs, minimizing inter-tile communication bandwidth by only forwarding condensed tile-level outputs to the chip-level accumulation unit for final cross-tile accumulation.

%===============================================================================
% V. IMPLEMENTATION
%===============================================================================
\section{Implementation Details}
\label{sec:implementation}

\subsection{TransCIM Framework}
\label{subsec:transcim}

We introduce TransCIM (Transformer CIM), a simulation framework built on the backbone of the open-source NeuroSim~\cite{peng2019neurosim} platform, to enable end-to-end evaluation of transformer workloads on CIM architectures. TransCIM extends NeuroSim's baseline two-operand CIM models to accommodate all transformer layer types: attention stages (including the proposed trilinear operations), linear projections (MLP layers, embedding matrices), and digital operations (Softmax, LayerNorm, GELU). In this sense, TransCIM is an experimentally calibrated simulation framework: it inherits NeuroSim's validated circuit models and established architecture/floorplanning backbone for CMOS peripherals, ADCs, buffers, interconnects, and standard FeFET read/write paths~\cite{peng2019neurosim,peng2020neurosimv2,chen2018neurosim}, and augments this backbone with a calibrated DG-FeFET device-to-crossbar model that captures non-volatile top-gate weight storage, volatile back-gate operand modulation, and the associated readout/periphery costs using published DG-FeFET measurements~\cite{jiang2025bio}. The proposed crossbar behavior is evaluated within TransCIM's integrated device-to-architecture modeling flow.

The framework operates in two execution modes. The \emph{digital baseline} mode quantizes inputs and weights to INT8 but accumulates in FP32 with no ADC or output quantization, serving as a quantization-aware accuracy ceiling. The \emph{CIM emulation} mode adds hardware-aware effects including ADC quantization (output clipped to ADC bit-width), the bounded $\eta_{BG}$ approximation over the selected DG-FeFET conductance range (Section~\ref{subsec:trilinear}), and the hierarchical adder-based accumulation path described in Sections~\ref{subsec:crossbar} and~\ref{subsec:digital_ops}. Both modes apply identical INT8 input/weight quantization, isolating the accuracy impact of analog non-idealities from the quantization baseline.

INT8 quantization parameters are obtained via post-training quantization (PTQ): activation scales are calibrated on a small representative dataset, after which weights and inputs are quantized using a symmetric uniform scheme. Multi-bit weights are mapped to multiple cells when device precision is limited---for example, an 8-bit weight with 2-bit cells uses 4 adjacent cells per synapse, with a shift-add stage recombining partial sums ($\text{output} = \sum_i \text{partial}_i \times 2^{i \cdot b_\text{cell}}$), where $b_\text{cell}$ is the number of bits stored per cell. Input voltages are applied bit-serially via the switch matrix, cycling from LSB to MSB over multiple time steps.
% Note: For the 1st operand (WL input), DAC is not modeled in PPA---only bit-serial switch matrix.
% Back-gate (3rd operand) has two options: DAC-based (default) or bit-serial switch matrix.

\subsection{TransCIM PPA Modeling}
\label{subsec:transcim_ppa}

TransCIM derives its performance, power, and area (PPA) estimates from the validated circuit models of the underlying NeuroSim backbone~\cite{peng2019neurosim}. We adopt a heterogeneous integration approach: CMOS peripheral circuits (ADCs, multiplexers, sense amplifiers, shift-add logic, buffers, and drivers) are modeled at a 7nm FinFET technology node using TSMC/IRDS transistor parameters~\cite{peng2019neurosim}, while FeFET memory cells use 22nm device characteristics (write voltage = 4.0V, write pulse = 50ns, $R_\text{on}$ = 240k$\Omega$, $R_\text{off}$ = 24M$\Omega$), consistent with the ferroelectric top-gate stack used in both single-gate~\cite{ni2018fefet} and DG-FeFET~\cite{jiang2025bio} implementations. This can be extended to a back-end-of-line (BEOL) integration where NVM devices are fabricated at a relaxed feature size above the dense CMOS logic layer---a physically realistic configuration since FeFET devices do not scale as aggressively as CMOS transistors~\cite{ni2018fefet}. Conductance values are constrained within the linear operating band established in Section~\ref{subsec:trilinear}, which is calibrated to experimental device data~\cite{jiang2025bio}.

% Token chunking technical note (for internal reference):
% chunk_attention_full_sequence=True ensures cross-chunk (all-to-all) attention.
% Chunking is purely a hardware parallelism technique for PPA scaling, not algorithmic.

\begin{table}[t]
    \centering
    \caption{Default system configuration.}
    \label{tab:system_config}
    \begin{tabular}{ll}
    \toprule
    \textbf{Parameter} & \textbf{Value} \\
    \midrule
    Technology node & 7nm CMOS / 22nm FeFET \\
    SubArray size & 64$\times$64 (scalable) \\
    Input precision & 8-bit \\
    Weight precision & 8-bit (2-bit/cell) \\
    ADC precision & 8-bit \\
    Column muxing & 8:1 (ADC sharing) \\
    Back-gate operand & 8-bit input, 0--5 V range \\
    Write voltage & 4.0 V~\cite{ni2018fefet} \\
    Write pulse & 50 ns~\cite{ni2018fefet} \\
    Global buffer & 4MB SRAM\textsuperscript{*} \\
    \bottomrule
    \multicolumn{2}{l}{\footnotesize \textsuperscript{*} Scales linearly with sequence length (4MB valid for seq.\ length 64).}
    \end{tabular}
\end{table}

For trilinear operations, the back-gate modulation energy model accounts for: (1)~device back-gate capacitance, (2)~wire capacitance along the per-column back-gate lines (estimated at 0.2~fF/$\mu$m for local interconnect), (3)~driver circuits that buffer and distribute the signal, and (4)~level shifting for the 0--5~V back-gate range. These components contribute to the core read energy of Stages~2 and~3, where the back-gate operand varies with the token. In the default 64$\times$64 SubArray configuration, the modeled back-gate modulation cost is 34.11~pJ per SubArray query step and remains below 1.2\% of trilinear core dynamic energy. The DC offset term (the ``1'' in Eq.~\ref{eq:trilinear_primitive}) is removed through a reference read with $V_{BG}=0$ on the same crossbar under the same row-input condition, which measures the $V_D\cdot G_0$ component for subtraction from the trilinear readout. Reference/bias compensation is included in the modeled PPA accounting and remains below 1\% of per inference energy in the trilinear mode.
Multi-head attention parallelism is modeled with latency taking the maximum across parallel heads and energy summing across all heads, reflecting the hardware reality of concurrent but independent head computation. Table~\ref{tab:system_config} lists the default hardware parameters used unless otherwise noted in Section~\ref{sec:experiments}.

%===============================================================================
% VI. EXPERIMENTS
%===============================================================================
\section{Experiments}
\label{sec:experiments}

\subsection{Experimental Setup}
\label{subsec:setup}

We evaluate the trilinear CIM accelerator on two representative transformer architectures: BERT-base-uncased~\cite{devlin2019bert} (12 layers, 12 heads, $d$=768) for NLP and ViT-base~\cite{dosovitskiy2020vit} (12 layers, 12 heads, $d$=768) for computer vision.

For NLP evaluation, we use nine tasks from the GLUE benchmark~\cite{wang2018glue}: CoLA, SST-2, MRPC, RTE, STS-B, WNLI, QNLI, QQP, and MNLI, covering sentiment analysis, paraphrase detection, textual entailment, and semantic similarity. For vision, we evaluate on CIFAR-10, CIFAR-100~\cite{krizhevsky2009cifar}, and ImageNet-1K~\cite{deng2009imagenet}. All models use INT8 post-training quantization for both weights and activations.

We compare three evaluation modes.
\textbf{Quantized-Digital} performs inference on ideal digital hardware with no analog non-idealities, serving as the accuracy ceiling.
\textbf{CIM-Bilinear} uses conventional single-gate FeFET CIM where $K$/$V$ matrices are dynamically reprogrammed onto the crossbar each sequence, subject to write latency overhead.
\textbf{CIM-Trilinear} is the proposed DG-FeFET architecture where projection weights remain static on the top gate and the input sequence drives sequence-dependent modulation of conductance via the back-gate.

For the controlled bilinear-vs.-trilinear comparison, both CIM modes are evaluated in TransCIM with the same model checkpoints, PTQ calibration, input/weight precisions, ADC/MUX settings, CMOS peripheral models, FeFET read/write parameters, and floorplanning procedure. The matched baseline therefore isolates the architectural change: dynamic attention operands are either rewritten into FeFET arrays (CIM-Bilinear) or delivered through the volatile DG-FeFET back-gate (CIM-Trilinear). Reported GPU and prior-accelerator numbers are included only for qualitative positioning, not as normalized comparisons.

\subsection{Accuracy Results}
\label{subsec:accuracy}

Tables~\ref{tab:glue_accuracy} and~\ref{tab:vision_accuracy} summarize accuracy across NLP and vision benchmarks, respectively. All reported values are mean $\pm$ standard deviation over three independent runs per evaluation mode.

\textbf{GLUE Results (Table~\ref{tab:glue_accuracy}).}
The trilinear mode outperforms bilinear on seven of nine tasks, with the largest improvements on QNLI (+3.74\%), MNLI (+3.14\%), STS-B (+1.74\%), and SST-2 (+1.60\%). This consistent advantage arises because bilinear repeatedly forms intermediate tensors across mixed-signal interfaces: analog CIM outputs must be digitized, requantized/remapped, and written back to the array before subsequent operations. These extra conversion and remapping steps accumulate numerical error, whereas trilinear keeps all projection weights static and applies the dynamic operand through the volatile back-gate. 
The lower standard deviation of trilinear results (typically $<$1\%) compared to bilinear (up to ${\sim}8.5$\%) further suggests that eliminating these repeated mixed-signal conversion and remapping steps yields more stable inference. The only task where trilinear underperforms bilinear is RTE ($-$1.33\%). %RTE's small validation set amplifies the impact of analog noise, with each flipped prediction shifting accuracy by ${\sim}$0.36\%. Combined with textual entailment's reliance on fine-grained premise--hypothesis token interactions, this likely makes RTE more susceptible to the additional back-gate quantization stage than coarser classification tasks. 
CoLA shows a modest trilinear advantage ($+$1.53\%), but both CIM modes remain well below digital ($-$1.76 and $-$3.29 for trilinear and bilinear, respectively). This gap likely reflects the Matthews Correlation Coefficient's heightened sensitivity to false-positive/negative balance under analog noise. WNLI yields identical results across all three modes (56.34\%) because its tiny dataset converges to the majority-class baseline regardless of hardware non-idealities.

\textbf{Vision Results (Table~\ref{tab:vision_accuracy}).}
In contrast to GLUE, the gap between trilinear and digital widens from CIFAR-10 to CIFAR-100 to ImageNet, while bilinear remains consistently closer to digital across all three ViT benchmarks. This reversal suggests that, for ViT, the error introduced by the back-gate quantization path outweighs the accuracy benefit obtained by avoiding bilinear's repeated mixed-signal conversion and remapping steps.

The trilinear computation introduces an additional quantization layer via the back-gate DAC that bilinear does not require: the dynamic modulating operand is discretized through a uniform DAC before it modulates device conductance. For NLP tasks, this extra quantization is well-tolerated because discrete token semantics provide natural noise resilience---small perturbations to attention weights rarely change which token is attended. For ViT, however, attention maps exhibit extreme non-uniform distributions with sparse, high-magnitude outlier scores critical for patch discrimination~\cite{lin2022fqvit, yuan2022ptq4vit}. The uniform back-gate DAC systematically distorts these outlier values, collapsing the sharp attention peaks that ViT relies on. Additionally, ViT post-LayerNorm activations exhibit wider inter-channel variance than NLP models~\cite{li2023repqvit}, further amplifying the back-gate quantization error.

\begin{table}[t]
\caption{GLUE benchmark scores. Bold indicates the best CIM mode per task. Seq: sequence length in tokens. Metrics---MCC: Matthews Correlation Coefficient; Acc: Accuracy; F1: F1-score; Pears.: Pearson correlation.}
\label{tab:glue_accuracy}
\centering
\begin{tabular}{llccc}
\toprule
Task (Metric) & Seq & Digital & Bilinear & Trilinear \\
\midrule
CoLA (MCC)     & 64  & 45.46{\tiny$\pm$1.46} & 42.17{\tiny$\pm$8.49}  & \textbf{43.70}{\tiny$\pm$1.03}  \\
SST-2 (Acc)    & 64  & 91.59{\tiny$\pm$0.40} & 89.72{\tiny$\pm$1.07}  & \textbf{91.32}{\tiny$\pm$0.43}  \\
MRPC (F1)      & 128 & 88.04{\tiny$\pm$0.87} & 84.24{\tiny$\pm$3.62}  & \textbf{85.54}{\tiny$\pm$0.62}  \\
RTE (Acc)      & 128 & 70.40{\tiny$\pm$0.72} & \textbf{68.11}{\tiny$\pm$2.18}  & 66.78{\tiny$\pm$1.53}  \\
STS-B (Pears.) & 128 & 85.79{\tiny$\pm$0.23} & 82.02{\tiny$\pm$2.63}  & \textbf{83.76}{\tiny$\pm$0.77}  \\
WNLI (Acc)     & 128 & 56.34{\tiny$\pm$0.00} & \textbf{56.34}{\tiny$\pm$0.00}  & \textbf{56.34}{\tiny$\pm$0.00}  \\
QNLI (Acc)     & 128 & 88.21{\tiny$\pm$0.08} & 82.04{\tiny$\pm$4.79}  & \textbf{85.78}{\tiny$\pm$0.93}  \\
QQP (F1)       & 128 & 86.53{\tiny$\pm$0.24} & 82.61{\tiny$\pm$2.93}  & \textbf{83.45}{\tiny$\pm$0.30}  \\
MNLI (Acc)     & 128 & 75.67{\tiny$\pm$0.35} & 72.44{\tiny$\pm$3.63}  & \textbf{75.58}{\tiny$\pm$0.55}  \\
\bottomrule
\end{tabular}
\end{table}

\begin{table}[t]
\caption{Vision benchmark accuracy (\%) for ViT-base (default configuration per Table~\ref{tab:system_config}).}
\label{tab:vision_accuracy}
\centering
\begin{tabular}{lccc}
\toprule
Dataset       & Digital & Bilinear & Trilinear \\
\midrule
CIFAR-10      & 97.01{\tiny$\pm$0.28} & 96.74{\tiny$\pm$0.65}  & 95.53{\tiny$\pm$0.09}  \\
CIFAR-100     & 86.05{\tiny$\pm$0.28} & 83.90{\tiny$\pm$0.84}  & 81.44{\tiny$\pm$1.40}  \\
ImageNet-1K   & 79.39{\tiny$\pm$1.04} & 79.11{\tiny$\pm$1.06}  & 74.98{\tiny$\pm$2.05}  \\
\bottomrule
\multicolumn{4}{l}{\footnotesize ViT processes 197 tokens per image.}
\end{tabular}
\end{table}

\subsection{PPA Analysis}
\label{subsec:ppa}

Table~\ref{tab:floorplan_summary} first summarizes the floorplanned hardware scale for the default 64$\times$64 SubArray, 2-bit/cell and 8-bit-ADC (2b/8b) evaluations using the NeuroSim-based floorplanning flow~\cite{peng2019neurosim}. BERT-base and ViT-base have similar tile counts because both use base-transformer dimensions and 64-token chunked execution in TransCIM (i.e., physical tiles are sized for 64-token chunks). Table~\ref{tab:ppa_seqlen} then presents the per-inference performance, power, and area (PPA) comparison between bilinear and trilinear for BERT-base inference. Because per-inference PPA is dominated by model architecture and sequence length rather than dataset content, tasks sharing the same sequence length exhibit only minor differences in these metrics.

\begin{table}[t]
\centering
\caption{Evaluated 64$\times$64 hierarchy and floorplan summary for 2b/8b runs.}
\label{tab:floorplan_summary}
\setlength{\tabcolsep}{3pt}
\begin{tabular}{lrrrrr}
\toprule
\textbf{Config} & \textbf{Bil. Tiles} & \textbf{Tri. Tiles} & \textbf{Attn.\ Tiles} & \textbf{Bil. SAs} & \textbf{Tri. SAs} \\
\midrule
BERT & 1{,}886 & 2{,}606 & 1{,}440 & 120{,}704 & 166{,}784 \\
ViT  & 1{,}900 & 2{,}620 & 1{,}440 & 121{,}600 & 167{,}680 \\
\bottomrule
\end{tabular}
\par\vspace{1pt}
\parbox{0.95\columnwidth}{\footnotesize\emph{Note:} Bil./Tri. denote CIM-Bilinear/CIM-Trilinear; Attn.\ Tiles count attention-stage tiles in the CIM-Trilinear mode; SAs denote SubArrays.}
\end{table}

\textbf{Per-Inference Efficiency (Table~\ref{tab:ppa_seqlen}).}
The trilinear mode achieves consistent improvements at both evaluated sequence lengths (64 and 128 tokens): 18.6--20.4\% latency reduction and 39.7--46.6\% energy savings, with the largest energy gains on 64-token tasks. In TransCIM, the energy saved by eliminating dynamic writes becomes less significant at longer sequence lengths. This trend reflects the different scaling of the underlying operations. As sequence length increases, attention computation must evaluate interactions between many more token pairs, so the associated CIM reads and accumulations grow approximately quadratically. In contrast, bilinear write energy is tied to forming dynamic operands once per token and therefore grows only linearly. Consequently, at sequence length~128, the energy saved by eliminating dynamic writes becomes a smaller fraction of total inference energy. Trilinear energy efficiency reaches 12.2--13.5~TOPS/W under default GLUE sequence-length caps (64 tokens for CoLA/SST-2 and 128 tokens for the remaining tasks), an improvement of 23--39\% over bilinear at the same caps, demonstrating that eliminating write operations yields substantial efficiency gains beyond the raw energy savings.

\textbf{Area and Compute Density.}
The trilinear mode incurs a constant 37.3\% area overhead, independent of sequence length, reflecting the additional back-gate delivery circuitry. Compute density is mixed: at 128-token context, TOPS/mm$^2$ increases by 9.5\% versus bilinear, whereas at 64-token context the smaller bilinear footprint yields ${\sim}$10\% \emph{lower} trilinear TOPS/mm$^2$ despite higher trilinear TOPS/W. Memory utilization remains stable at ${\sim}$87\% for trilinear (vs.\ ${\sim}$84\% for bilinear), reflecting slightly better tile-level packing under the trilinear attention mapping.

\begin{table}[t]
\caption{Per-inference PPA comparison by sequence length (BERT-base, default configuration per Table~\ref{tab:system_config}).}
\label{tab:ppa_seqlen}
\centering
\setlength{\tabcolsep}{3pt}
\begin{tabular}{lrrrrrrr}
\toprule
 & \multicolumn{3}{c}{\textbf{Seq 64}} & \multicolumn{3}{c}{\textbf{Seq 128}} \\
\cmidrule(lr){2-4} \cmidrule(lr){5-7}
\textbf{Metric} & \textbf{Bil.} & \textbf{Tri.} & \textbf{$\Delta$\%} & \textbf{Bil.} & \textbf{Tri.} & \textbf{$\Delta$\%} \\
\midrule
Area (mm$^2$)      & 326   & 447   & +37.3 & 651   & 894   & +37.3 \\
Latency (ms)       & 7.63  & 6.08  & $-$20.4 & 8.19  & 6.67  & $-$18.6 \\
Energy ($\mu$J)    & 1{,}522 & 813 & $-$46.6 & 3{,}132 & 1{,}889 & $-$39.7 \\
Throughput (inf/s) & 131   & 164   & +25.5 & 122   & 150   & +22.7 \\
TOPS/W             & 9.97  & 12.24  & +22.8 & 9.68  & 13.47  & +39.2 \\
TOPS/mm$^2$        & 0.0056  & 0.0050  & $-$9.9 & 0.0053  & 0.0058  & +9.5 \\
Mem.\ Util.\ (\%)  & 84.5  & 87.4  & +3.4  & 84.5  & 87.4  & +3.4  \\
\bottomrule
\end{tabular}
\end{table}

\begin{table}[t]
\caption{Per-inference component breakdown for BERT-base at sequence length 128 under the default 2b/8b, 64$\times$64-SubArray configuration.}
\label{tab:component_breakdown}
\centering
\setlength{\tabcolsep}{3pt}
\begin{tabular}{lrr}
\toprule
\textbf{Component} & \textbf{Bilinear} & \textbf{Trilinear} \\
\midrule
Core dynamic energy ($\mu$J) & 1{,}501 & 1{,}537 \\
Support dynamic energy ($\mu$J) & 1{,}189 & 69 \\
Leakage energy ($\mu$J) & 437 & 282 \\
ADC share of core dynamic energy & $\sim$83\% & $\sim$82\% \\
Accum.\ share of core dynamic energy & $\sim$3.9\% & $\sim$3.8\% \\
ADC/sense latency ($\mu$s) & 752 & 574 \\
Accum.\ latency ($\mu$s) & 248 & 176 \\
Other latency ($\mu$s) & 5{,}526 & 5{,}922 \\
Attention-write latency ($\mu$s) & 1{,}376 & 0 \\
Total latency (ms) & 8.19 & 6.67 \\
Attention-write energy ($\mu$J) & 49.7 & 0 \\
$V_{BG}$ delivery cost & -- & 34.11~pJ/step \\
Reference/bias comp.\ & -- & $<$1\% total \\
\bottomrule
\end{tabular}
\end{table}

The component breakdown in Table~\ref{tab:component_breakdown} shows that core dynamic energy is comparable between modes because trilinear uses additional active tiles for fused three-operand attention compute. The dominant saving comes from support dynamic energy: buffer, interconnect, and DRAM traffic associated with dynamic $K/V$ reprogramming drops by 94\%, while attention-write latency and energy are eliminated. Table~\ref{tab:vit_ppa} extends the PPA comparison to ViT-base with 197 tokens/image, showing that the trilinear mapping also improves energy efficiency for vision workloads.

\begin{table}[t]
\caption{ViT-base PPA with 197 tokens/image and 2b/8b precision; 64$\times$64 and 32$\times$32 denote SubArray sizes.}
\label{tab:vit_ppa}
\centering
\setlength{\tabcolsep}{3pt}
\begin{tabular}{lrrrr}
\toprule
 & \multicolumn{2}{c}{\textbf{64$\times$64}} & \multicolumn{2}{c}{\textbf{32$\times$32}} \\
\cmidrule(lr){2-3} \cmidrule(lr){4-5}
\textbf{Metric} & \textbf{Bil.$\to$Tri.} & \textbf{$\Delta$} & \textbf{Bil.$\to$Tri.} & \textbf{$\Delta$} \\
\midrule
Area (mm$^2$) & 1{,}000$\to$1{,}374 & +37.4\% & 1{,}581$\to$1{,}862 & +17.8\% \\
Latency (ms) & 8.96$\to$7.57 & $-$15.6\% & 10.31$\to$6.44 & $-$37.6\% \\
Energy ($\mu$J) & 4{,}946$\to$3{,}273 & $-$33.8\% & 6{,}657$\to$5{,}094 & $-$23.5\% \\
TOPS/W & 9.48$\to$14.72 & +55.3\% & 6.52$\to$10.08 & +54.6\% \\
\bottomrule
\end{tabular}
\end{table}

\subsection{Ablation Studies}
\label{subsec:ablation}

\begin{figure}[t]
\centering
\includegraphics[width=\columnwidth]{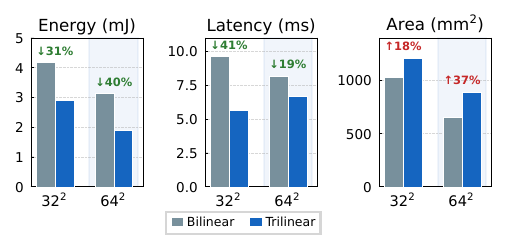}
\caption{SubArray size ablation (2b/8b, seq\,=\,128, per inference). Energy, latency, and chip area for 32$^2$ and 64$^2$ SubArrays. Green annotations denote trilinear improvement over bilinear; red annotations denote area overhead. The 64$^2$ default (shaded) is used throughout the paper.}
\label{fig:ablation_panels}
\end{figure}

\begin{figure}[t]
\centering
\includegraphics[width=\columnwidth]{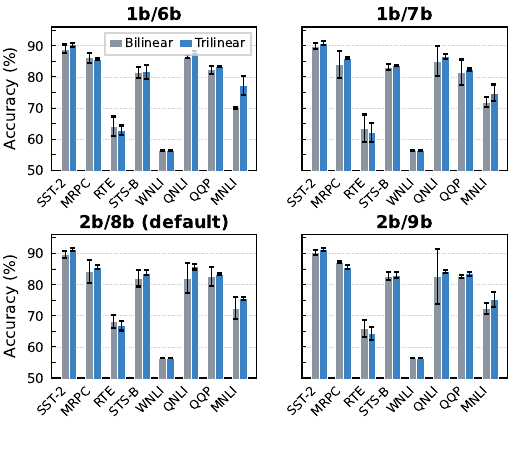}
\caption{Per-task GLUE benchmark scores for four bitcell/ADC configurations (SA=64$\times$64, seq\,=\,128). Grey: bilinear; blue: trilinear. Error bars show $\pm$1 std over 2--3 seeds. CoLA (Matthews Corr.) omitted due to its different metric scale.}
\label{fig:precision_accuracy}
\end{figure}

We ablate three design-space axes---SubArray size, bitcell/ADC precision, and sequence length---to characterize the trilinear architecture's sensitivity to key parameters. Fig.~\ref{fig:ablation_panels} visualizes the SubArray size sweep, while Table~\ref{tab:ablation_precision} and Fig.~\ref{fig:precision_accuracy} detail the precision sweep's PPA and per-task performance metric, respectively.

\textbf{A. SubArray Size (2b/8b).}
Fig.~\ref{fig:ablation_panels} compares 32$\times$32 and 64$\times$64 SubArrays with bitcell (2b) and ADC precision (8b) fixed; PPA values are per-inference metrics representative of all seq\,=\,128 GLUE tasks.
As the figure shows, 32$^2$ yields the largest latency reduction ($-$40.9\%) and near-perfect memory utilization, while 64$^2$ delivers the larger relative energy reduction on this seq\,=\,128 bucket ($-$39.7\% vs.\ $-$30.9\%) and the higher absolute TOPS/W (13.47 vs.\ 9.38). Although the trilinear area \emph{overhead} is smaller at 32$^2$ (+17.8\% vs.\ +37.3\%), its absolute chip area remains larger because smaller SubArrays replicate more peripheral circuitry. TOPS/W improves at both sizes (6.70$\to$9.38 at 32$^2$; 9.68$\to$13.47 at 64$^2$), and the higher absolute TOPS/W at 64$^2$ reflects its greater per-read analog parallelism (64 rows vs.\ 32).
Accuracy remains robust at 32$^2$: trilinear SST-2 reaches 91.70$\pm$0.05\% (vs.\ bilinear 89.64$\pm$0.52\%) and MNLI 73.48$\pm$0.70\% (vs.\ 67.77$\pm$6.46\%), closely tracking or matching the 64$^2$ default (Table~\ref{tab:glue_accuracy}). The substantially lower trilinear standard deviation at 32$^2$ ($<$1\% on most tasks) versus bilinear (up to ${\sim}$6.5\%) reinforces the stability advantage of write-free inference.
Across both configurations, trilinear avoids dynamic writes entirely (0 vs.\ 18.9M physical cell writes per inference for bilinear at seq\,=\,128).

\textbf{B. Bitcell/ADC Precision (SA=64$\times$64).}
Table~\ref{tab:ablation_precision} summarizes PPA for four bitcell/ADC configurations, with per-task performance metric detailed in Fig.~\ref{fig:precision_accuracy}.
The 1b/6b configuration emerges as the strongest accuracy point: trilinear achieves the largest per-task improvements over bilinear (MNLI $+$7.13\%, SST-2 $+$1.30\% as shown in Fig.~\ref{fig:precision_accuracy}, and CoLA $+$9.66\%), while delivering 37.5\% energy reduction and 26.0\% latency improvement at only 32.4\% area overhead---consistently lower than the ${\sim}$37\% incurred by 2-bit configurations (Table~\ref{tab:ablation_precision}), making it particularly compelling for accuracy-critical deployments.

The 2b/7b configuration (omitted from the table) demonstrates that 2-bit cells require at least 8-bit ADC precision---below this threshold, both CIM modes collapse to chance-level accuracy, establishing ADC precision as the binding constraint for multi-bit cells.
Increasing ADC precision beyond the default brings diminishing returns: 2b/9b gains only marginal accuracy while increasing energy and area (37.4\% area overhead). Similarly, 1b/7b and 1b/8b (not shown) offer progressively lower TOPS/W, confirming that 1b/6b is the optimal low-precision operating point.

\begin{table}[t]
\caption{Bitcell/ADC precision ablation (SA=64$\times$64, seq~128). PPA deltas are trilinear vs.\ bilinear. Per-task GLUE score (task-specific metric; see Table~\ref{tab:glue_accuracy}) is shown in Fig.~\ref{fig:precision_accuracy}.}
\label{tab:ablation_precision}
\centering
\begin{tabular}{lrrrrrr}
\toprule
\textbf{Config} & \textbf{Area} & \textbf{Lat.} & \textbf{Energy} & \multicolumn{2}{c}{\textbf{TOPS/W}} \\
 & \textbf{$\Delta$\%} & \textbf{$\Delta$\%} & \textbf{$\Delta$\%} & \textbf{Bil.} & \textbf{Tri.} \\
\midrule
1b/6b  & +32.4 & $-$26.0 & $-$37.5 & 10.45 & 13.77 \\
1b/7b  & +32.4 & $-$35.4 & $-$32.5 & 7.52 & 10.22 \\
%1b/8b  & +32.5 & $-$32.4 & $-$25.7 & 4.70 & 6.38 \\
%2b/7b  & +37.2 & $-$20.5 & $-$50.7 & 17.32 & 23.43 \\
2b/8b  & +37.3 & $-$18.6 & $-$39.7 & 9.68 & 13.47 \\
2b/9b  & +37.4 & $-$27.5 & $-$31.5 & 5.31 & 7.61 \\
\bottomrule
\end{tabular}
\end{table}

\textbf{C. Sequence Length Scaling (2b/8b, SA=64$\times$64).}
The trilinear advantage in energy and latency is maintained when each GLUE task's default context is doubled: CoLA and SST-2 scale from 64 to 128 tokens, while MRPC, RTE, STS-B, WNLI, QNLI, QQP, and MNLI scale from 128 to 256 tokens. For CoLA and SST-2, doubling context from 64 to 128 tokens changes the energy reduction from 46.6\% to 39.7\% and the latency reduction from 20.4\% to 18.6\%, while the TOPS/W gain rises from +22.8\% to +39.2\%. For the remaining GLUE tasks, doubling context from 128 to 256 tokens changes the energy reduction from 39.7\% to 27.4\% and the latency reduction from 18.6\% to 16.1\%, while the TOPS/W gain rises from +39.2\% to +68.6\%. Accuracy remains task-dependent under doubled context.

From an endurance perspective, the critical advantage at longer sequences appears in write volume: in the doubled-context sweep, bilinear incurs 18.9M cell writes per inference for the 128-token CoLA/SST-2 bucket and 37.7M for the 256-token bucket used by the remaining GLUE tasks, while trilinear remains at exactly zero. These doubled-context sweeps extend the evaluated full-sequence setting from the GLUE defaults (64/128 tokens) to 128/256 tokens, rather than claiming coverage of modern data-center LLM contexts; within this measured range, TrilinearCIM preserves zero runtime attention writes as context length increases.

\subsection{Discussion}
\label{subsec:discussion}

\textbf{Design trade-offs.}
The ablation studies reveal two promising operating points beyond the default: (1)~the 1b/6b configuration, which delivers the strongest accuracy advantage while still reducing energy by 37.5\% at only 32.4\% area overhead (Table~\ref{tab:ablation_precision}, Fig.~\ref{fig:precision_accuracy}); (2)~the 32$\times$32 SubArray, which delivers the largest latency reduction, near-perfect memory utilization, and a smaller trilinear-vs.-bilinear area overhead than 64$^2$ (+17.8\% vs.\ +37.3\%) (Fig.~\ref{fig:ablation_panels}). The default 64$\times$64 with 2b/8b balances these trade-offs by combining reliable accuracy across all GLUE tasks with higher absolute TOPS/W than the 32$\times$32 alternative and a smaller absolute chip footprint on the seq\,=\,128 bucket.

\textbf{Area overhead justification.}
The area overhead (Table~\ref{tab:ppa_seqlen}) is partially offset at 128-token context by improved TOPS/mm$^2$; at 64-token context, energy efficiency gains dominate despite slightly lower trilinear TOPS/mm$^2$. These results suggest that the extra area can be justified when energy savings are prioritized. %The area overhead (Table~\ref{tab:ppa_seqlen}) is partially offset by improved compute density: TOPS/mm$^2$ increases at both sequence lengths, indicating that the added back-gate driver circuitry is used effectively. These results suggest that the extra area can be justified when energy savings and compute density are prioritized.

\textbf{Scalability.}
The sequence length sweep (Section~\ref{subsec:ablation}) confirms that the trilinear advantage is maintained for the evaluated context-length increase from 64/128 tokens to 128/256 tokens. TrilinearCIM is positioned for edge-AI attention workloads evaluated as complete input sequences, including encoder-style models and decoder prefill/full-sequence causal attention. The current design and measurements do not implement token-by-token autoregressive KV-cache decoding; extending the dataflow to incremental LLM decoding and persistent KV-cache management is outside the scope of this evaluation and is left for future work. Scaling dense attention to thousand-token contexts would require sparse or linear-attention algorithms orthogonal to the CIM dataflow that eliminates runtime NVM reprogramming of dynamic attention operands. For Mixture-of-Experts (MoE) models, resident FFN experts can be selected by routing active DG-FeFET arrays, and attention-expert variants can reuse the same VMVM schedule; expert placement beyond on-chip capacity is a higher-level scheduling problem rather than a change to the MAC primitive. Like prior NVM-CIM transformer literature~\cite{sridharan2023xformer,yang2020retransformer,laguna2022hardware,zhou2022transpim}, our empirical evaluation targets encoder-style workloads, consistent with this community's scope.

\textbf{Limitations.}
Our evaluation relies on TransCIM (Section~\ref{sec:implementation}), which includes ADC quantization and uses a band-averaged $\eta_{BG}$ over the selected $G_0 \in [29,69]~\mu$S DG-FeFET conductance range. This range bounds the residual back-gate-sensitivity approximation error, while the evaluated SubArray partitioning follows common CIM practice for managing parasitic voltage-drop effects and peripheral tradeoffs~\cite{shafiee2016isaac,ankit2019puma,peng2019neurosim}. Prior FeFET characterization studies retention, temperature sensitivity, and device variation effects~\cite{saha2024fefet}, and prior FeFET IMC work shows that device-variation effects can be effectively mitigated through variation-aware probabilistic training~\cite{manna2024variation}. Full silicon characterization of a fabricated DG-FeFET trilinear array remains future work. We also leave algorithmic mitigation of the ViT accuracy gap to future work, including non-uniform back-gate quantization, hardware-aware fine-tuning, and quantization- or variation-aware training~\cite{li2023repqvit,manna2024variation}.

\textbf{Endurance.}
Under our operating assumption, trilinear attention computation bypasses ferroelectric switching: back-gate modulation acts through a non-ferroelectric dielectric and therefore does not intentionally perturb the stored polarization state. Read-disturb and other device-level reliability effects should still be validated experimentally.

%===============================================================================
% VII. RELATED WORK
%===============================================================================
\section{Related Work}
\label{sec:related_work}

\begin{table*}[t]
\centering
\caption{Comparison of attention-computation strategies across CIM, PIM, and digital transformer accelerators. Reported-PPA values are as published by each work; they are not directly comparable across technology nodes, devices, precision, or workloads.}
\label{tab:related_comparison}
\renewcommand{\arraystretch}{1.15}
\begin{tabular}{@{} p{1.5cm} p{1.9cm} p{4.5cm} p{1.0cm} p{2.6cm} p{4.7cm} @{}}
\toprule
\textbf{Work} & \textbf{Technology (Node)} & \textbf{Attention Strategy} & \textbf{NVM Writes} & \textbf{Reported PPA} & \textbf{Limitation} \\
\midrule
ReTransformer ~\cite{yang2020retransformer} & ReRAM (130\,nm) & Decomposes $QK^T{=}(QW_K)X^T$ to keep static $W_K$ on crossbar, avoiding $K^T$ write & Reduced & 467.68~GOPS/s/W & Score$\cdot V$ still writes $V$ to ReRAM; write endurance consumed \\
Sanger~\cite{lu2021sanger} & Digital ASIC (65\,nm) & Reconfigurable score-stationary architecture with structured attention pruning & --- & --- & Digital-only; no NVM weight integration \\
X-Former~\cite{sridharan2023xformer} & ReRAM + CMOS (65\,nm) & Dual-engine: ReRAM for static projections, CMOS attention engine for dynamic attention & Zero & 69.8$\times$ latency, 13$\times$ energy vs.\ GTX~1060 & Forfeits NVM density for the attention engine; CMOS area and leakage overhead \\
iMTransformer ~\cite{laguna2022hardware} & FeFET + CMOS (---) & FeFET crossbars for static projections; CMOS crossbars/CAMs cache attention scores & Zero & --- & Heterogeneous FeFET+CMOS fabric; attention engine sacrifices NVM density \\
TransPIM~\cite{zhou2022transpim} & HBM/DRAM-PIM (---) & HBM-based PIM with token-parallel digital dataflow & --- & 734~GOPS/s; 22.1$\times$--114.9$\times$ GPU speedup & Volatile DRAM: static leakage, refresh overhead, limited compute density vs.\ NVM \\
Yang et al.~\cite{yang2022fullcircuit} & Memristor (---) & Full-analog memristor circuit with capacitor-based analog memory for attention intermediates & Zero & --- & Capacitor charge decay limits retention; demonstrated only at $3{\times}3$ scale \\
\midrule
\textbf{Trilinear CIM (Ours)} & \textbf{DG-FeFET} \textbf{(22\,nm FeFET / 7\,nm CMOS)} & \textbf{Back-gate modulation enables in-memory attention without runtime ferroelectric rewriting} & \textbf{Zero} & \textbf{13.47~TOPS/W} & \textbf{Requires per-column BGL drivers and operation within the selected $G_0 \in [29, 69]\,\mu$S band} \\
\bottomrule
\end{tabular}
\end{table*}

\textbf{CIM Accelerators for Static Workloads.}
ISAAC~\cite{shafiee2016isaac} and PRIME~\cite{chi2016prime} pioneered ReRAM-based crossbar architectures with pipelined inter-layer execution and dual-mode memory-compute cells, respectively. PipeLayer~\cite{song2017pipelayer} extended this paradigm to on-chip training through weight-gradient parallelism. These designs target static weight matrices and do not address the dynamic operand challenge analyzed in Section~\ref{subsec:write_bottleneck}.

\textbf{Transformer-Specific CIM and Processing-in-Memory (PIM) Accelerators.}
Adapting CIM to Transformers requires addressing the dynamic nature of attention, where operands ($Q$, $K$, $V$) change with every input sequence. Prior works span four main strategies.
\textit{(i) Algorithmic write reduction:} ReTransformer~\cite{yang2020retransformer} reformulates the score computation as $Q \cdot K^T = (Q \cdot W_K) \cdot X^T$ via matrix decomposition, keeping the static projection weight $W_K$ on the ReRAM crossbar and avoiding the explicit write of the intermediate $K^T$ matrix. However, the value-aggregation stage ($\text{Score} \cdot V$) still requires writing $V$ to the crossbar, and the fundamental endurance and write-latency constraints of ReRAM remain.
\textit{(ii) Hybrid NVM+CMOS:} X-Former~\cite{sridharan2023xformer} partitions the workload across a ReRAM-based Projection Engine for static weights and a separate CMOS attention engine for the dynamic $Q \cdot K^T$ and $\text{Score} \cdot V$ operations, avoiding NVM writes entirely for attention at the cost of NVM density.
Similarly, iMTransformer~\cite{laguna2022hardware} combines FeFET crossbars for static projection weights with CMOS-based crossbars and CAMs that cache attention scores for reuse. The same group also explored FeFET-based attention-in-memory for few-shot learning~\cite{reis2021attention}.
\textit{(iii) DRAM-PIM:} TransPIM~\cite{zhou2022transpim} operates entirely in the DRAM domain using HBM-based processing-in-memory with a token-based dataflow, sidestepping NVM altogether but inheriting the limited compute density and leakage of volatile DRAM.
\textit{(iv) Full-analog transient storage:} Yang et al.~\cite{yang2022fullcircuit} propose a full-analog memristor circuit that stores intermediate attention results in capacitor-based analog memory rather than NVM, avoiding write endurance concerns but demonstrating only at a $3 \times 3$ scale.
Table~\ref{tab:related_comparison} summarizes these approaches and includes representative digital transformer accelerators for context.

\textbf{Digital Transformer Accelerators.}
In the purely digital domain, SpAtten~\cite{wang2021spatten} introduces cascade token and head pruning with progressive quantization to reduce attention computation and DRAM access.
A$^3$~\cite{ham2020a3} accelerates attention through approximate search, while Energon~\cite{zhou2022energon} employs mix-precision multi-round filtering to dynamically identify critical query-key pairs.
Sanger~\cite{lu2021sanger} co-designs structured attention pruning with a reconfigurable score-stationary architecture to reduce dense attention computation, operating entirely in the digital domain without NVM integration.
These designs achieve significant speedups over GPUs but remain fundamentally data-movement-bound, as the attention matrices must still traverse the memory hierarchy.

%===============================================================================
% VIII. CONCLUSION
%===============================================================================
\section{Conclusion}
\label{sec:conclusion}

This paper evaluates a DG-FeFET-based trilinear CIM dataflow that addresses the dynamic-operand bottleneck of Transformer attention without runtime ferroelectric rewriting of temporary attention operands. By using the back-gate as a third operand pathway, the design keeps projection weights stationary in the top gate while executing the attention MAC stages in NVM arrays within an end-to-end accelerator model that includes the surrounding digital operations.

Across BERT-base and ViT-base workloads, eliminating dynamic attention writes yields up to 20.4\% faster inference and 46.6\% lower energy per inference, with trilinear matching or exceeding the bilinear baseline on seven of nine GLUE tasks. The ablation sweep highlights 1b/6b as the accuracy-optimized point and 32$\times$32 as a latency-optimized, high-utilization alternative to the 64$\times$64 default. These benefits trade against additional back-gate circuitry and lower vision accuracy than the bilinear baseline under uniform back-gate quantization. Overall, the results show that trilinear DG-FeFET CIM can extend NVM-based acceleration beyond static projections to dynamic Transformer attention by replacing runtime ferroelectric reprogramming with volatile back-gate modulation.

\section*{Acknowledgments}
Research was sponsored primarily by the the National Science Foundation under award No. EFRI BRAID \#2318101. This work was also supported partially by the U.S. Department of Energy, Office of Science, Advanced Scientific Computing Research (ASCR) program as part of the NSF/NIH/DOE/ANR/BMBF/ BSF/NICT/AEI/ISCIII Collaborative Research in Computational Neuroscience (CRCNS) Program, project PARADIGM and the National Science Foundation under award No. ECCS \#2346953.

The authors acknowledge the use of LLMs in the preparation of this manuscript. Specifically, GPT-5.4 (OpenAI) and Claude 4.6 (Anthropic) were used for wording refinement, limited code and figure-script edits, and identifying potentially relevant prior work. All technical claims, numerical values, cited references, and final text were independently verified by the authors, who take full responsibility for the manuscript.

%%%%%%% -- PAPER CONTENT ENDS -- %%%%%%%%

%%
%% Bibliography
%%
\bibliographystyle{IEEEtran}
%\bibliography{references}

\end{document}